body of paper
\documentstyle[12pt]{article}
\begin{document}
\baselineskip24pt
\begin{center}
{\large\bf Arithmetic of the integer quantum Hall effect}
\vskip0.5cm
{\bf Vipin Srivastava}
\vskip0.25cm
{\it School of Physics, University of Hyderabad, Hyderabad - 500 046, India}
\end{center}
\vskip0.5cm
Integer quantum Hall effect (IQHE) has been analysed considering
the degeneracies of localized and extended states separately.
Occupied localized and extended states are counted and their
variation is studied as a function of magnetic field. The number
of current carrying electrons is found to have a saw-tooth
variation with magnetic field. The analysis attempts to answer
certain basic questions besides providing a simple but complete
understanding of IQHE.
\newpage
We show that in the integer-quantum-Hall setting the number of
current carrying electrons varies like saw-tooth with the
magnetic field. Infact we find that this is an alternative
manifestation of the integer-Hall-quantization.$^{1,2}$ We also
suggest an experiment for counting the number of extended and
localized states below the Fermi level as a function of magnetic
field $B$. Besides revealing some more interesting physics
embedded in the phenomenon of integer-quantum-Hall-effect (IQHE)
and providing the simplest way of understanding the fascinating
phenomenon, the present approach to the IQHE is expected to
resolve, through the suggested experiment, the following long
standing questions: (A) How does the IQHE approach the
2-dimensional localization result --- localization of all states
at any disorder$^3$ --- in the limit $B\to 0$? One has to resolve
between two apparently possible alternative scenarios, namely
(i) the extended states `float up' to infinite energy as $B\to
0;^4$ and (ii) the critical disorder $W_c$, required to localize
all states in a band approaches zero as $B\to 0$.$^5$ (B) Whether
the number of extended states in a Landau subband forms a
vanishing or a non-vanishing fraction of the total number of
states in the subband? We have also addressed two questions
related to the basic understanding to the IQHE: (C) How does the
IQHE acquire the spectacular accuracy and what are the factors
that put limit on it? And (D) How is it that exactly $ls(B)$
states ($s(B)$ being the degeneracy of a Landau subband) play the
central role in the integer Hall quantization$^6$ although all
$ls(B)$ states may not be occupied, or the number of occupied
states may far exceed $ls(B)$ for a value of $B$ at which the
Fermi level $E_F$ is located in the $l^{th}$ mobility gap?

We will count the number of extended and localized electrons as a
function of $B$, first assuming the Landau subbands to be
independent, and then by incorporating the result of Haldane and
Yang$^4$ to discuss the effect of band-mixing.

Take $B=0$ to start with and consider an increase in $B$ by
$\delta B$ that inserts {\it one} flux quantum into the system.
There will be $N$ Landau levels below $E_F$ in a system of $N$
electrons per unit area, and each level will have one state (spin
is not important for our purpose). Due to the presence of
disorder we ask: are the $N$ states (a) all localized; or (b) all
extended or (c) some localized and others extended? Neither (a)
nor (b) can hold {\it as a rule}, for then {\it all} subsequent
increments of $B$ by $\delta B$ would introduce either only
localized (in case (a)) or only extended states (in case (b)) and
consequently all states in the system would be either localized or
extended at all $B>0$. Both these possibilities are contrary to
the known results. Therefore, (c) must represent the true
situation. Now the question arises: as an increment $\delta B$
adds a new state to each Landau subband the fractions ofthe {\it new} 
states below $E_F$ that are respectively
localized and extended decided arbitrarily or is
there a rule governing it? We expect an underlying rule connected
with the fact that the amount of localization is decided by the
strength of disorder. So, for each Landau subband we should be
able to write,
\begin{equation}
\mbox{(no. of localized states)/(no. of extended states) = D},
\end{equation}
which besides depending on the strength of disorder should depend
on $B$ as well.

\noindent{\it The arithmetic:}  Recall that classically (without
disorder) the Hall voltage can be written as
\begin{equation}
{\cal E}_y(B) = s(B)vh/e\,\,\,\,,
\end{equation}
where $s(B)$ is the degeneracy of each Landau level and $v$ is
the average drift velocity of current carriers. In the presence
of disorder and localization we split $s(B)$ as
\begin{equation}
s(B) = s^E(B) + s^L(B)\,\,\,,
\end{equation}
with $E$ and $L$ respectively representing extended and localized
states, and write the Hall voltage in analogy with (2) as 
\begin{equation}
E_y(B) = s^E(B)V(B)h/e\,\,\,,
\end{equation}
keeping the system current density $j_x=n^E(B)eV(B)$ carried by
$n^E$ extended electrons unchanged at the value $Nev$ (as in a
typical IQHE experiment). The constancy of $j_x$ leads to
\begin{equation}
V(B) = (N/n^E(B))v\,\,\,,
\end{equation}
that is, $n^E$ electrons per unit area carry the current $Nev$ by
moving at a higher drift velocity $V$ to compensate for the loss
of current due to localization of $n^L$ $(=N-n^E)$ electrons.

The $s^E(B)$ in a particular band always increases with $B$
though non-monotonically --- it goes up by 1 only when
$\delta B$-increase of $B$ adds an extended state to this band
which happens with probability $1/(D+1)$ in view of (1) (note
that following (3), eqn.(1) will become $s^L(B)/s^E(B)=D$). But
we will see that $V(B)$ increases as well as decreases with $B$
depending on where $E_F$ is located. So, $E_y(B)$ can remain
unchanged with $B$ whenever $V(B)$ decreases, in case
\begin{equation}
s^E(B)V(B) = \mbox {a constant}\,\,\,\,.
\end{equation}
We will count the occupied localized and extended states as a
function of $B$ and investigate the quantum Hall plateaus through
(6) and address the questions stated above. We will follow the
picture of Fig.1 commonly used in connection with IQHE.$^7$

Suppose $E_F$ is located in the $l^{\rm th}$ mobility gap and the
numbers of occupied extended and localized states are
respectively $ls^E(B)$ and $ls^L(B)+\eta$ (see Fig.1d for
$\eta$), so that $V(B)=[\{ls(B)+\eta\}/\{ls^E(B)\}]v$. Now $B$ is
increased by $\delta B$, and $l$ new states --- one each in $l$
subbands below $E_F$ --- are added. Suppose $i$ of these states
are extended and $(l-i)$ localized. The $E_F$ will move downwards
by $l$ states and the numbers of extended and localized states
will become $(ls^E(B)+i)$ and $(ls^L(B)+\eta-i)$ respectively. Then,
\begin{equation}
s^E(B+\delta B)V(B+\delta B) = s^E(B+\delta B) {ls(B)+\eta\over
ls^E(B)+i} v = {s^E(B+\delta B)\over s^E(B)+i/l} s^E(B)V(B) \,\,\,.
\end{equation}
And, if $E_F$ lies in the $l^{\rm th}$  band of extended states,
then the counting of localized and extended states below $E_F$
would give
\begin{equation}
s^E(B+\delta B)V(B+\delta B) = {s^E(B+\delta B)\over
s^E(B)-(l-i)/l} s^E(B)V(B)\,\,\,\,.
\end{equation}
To get the behaviour of $E_y$ we will examine
$$
{\delta E_y(B)\over E_y} = {[s^E(B+\delta B)-s^E(B)]-i/l\over
s^E(B)+i/l} \,\,\mbox {for $ E_F$ in $l^{\rm th}$
mobility gap};\eqno(9a)
$$
and,
$$
= {1+[s^E(B+\delta B)-s^E(B)]-i/l\over
s^E(B)-(l-i)/l} \,\,\mbox {for $E_F$ in $l^{\rm th}$ band of ext.sts.};
\eqno(9b)
$$
\addtocounter{equation}{+1}

Note that $i$ can be 1 only with probability $l/(D+1)$, and that
$[s^E(B+\delta B)-s^E(B)]$ can only be either 0 or 1 for a
subband since $s^E$ should be an integer. So, in (9a) $\delta E_y$
remains zero until the magnetic field is incremented by
$[(D+1)/l]\delta B$ when $i$ becomes 1 with probability one and
the $s^E$ becomes $s^E(B)+1$ in one of the $l$ subbands, and
stays at value $s^E(B)$ in the
remaining $(l-1)$ subbands. The 
subband that gets the new extended state makes a non-zero
contribution to (9a). This makes $\delta E_y$ non-zero of order
$[10^5s(B)/(D+1)]^{-1}$.$^8$ The inaccuracy $\delta E_y$ remains
the same on the further increase of magnetic field until the next
extended state is introduced below $E_F$. In this way a plateau
is formed in the $E_y$ with an accuracy of few parts in $[10^6s(E)/(D+1)]$.

For the $V(B)$ note that when $E_F$ lies in a mobility gap we
will have either $V(B+\delta B)<V(B)$ whenever an extended state
is produced and the $s^E$ is enhanced in the subbands below
$E_F$, or $V(B+\delta B)=V(B)$ inbetween these events. On the
other hand when $E_F$ lies in a band of extended states we will
always have $V(B+\delta B)>V(B)$ because $n^E(B)$ will
necessarily decrease due to the downward movement of $E_F$. We
find here that good amount of information can be extracted from
the variation of $V(B)$ with $B$. Before we go into the details
of the variation of $V(B)$ we will understand the role played by
the flexibility of $V(B)$ in the light of the question (D).

If the given $N$ electrons {\it exactly} fill $l$ levels then
from (2), in the {\it classical case}
\begin{equation}
R_H = {{\cal E}_y(B)\over j_x} = {h\over le^2}\,\,\,,
\end{equation}
and this result can be maintained as independent of $N$ and $B$
{\it classically} by adding $l$ electrons to the system from
outside each time $B$ is increased by $\delta B$, and by
maintaining $j_x$ at $Nev$ (which reduces $v$ suitably as $N\to
N+l$). The IQHE presents a setting where the system, under
certain conditions, on its own mimics this classical scenario
--- quantum localization of electrons creates a buffer of states
which feeds electrons to $l$ completely filled bands of extended
electrons, and keeps them completely filled over a range of $B$.
As long as the bands of current - carrying electrons are exactly
filled and $j_x$ is maintained constant, the number of electrons
in the bands has no relevance, only the number of bands
matters for $R_H$ as in the above classical case. The exact 
filling of $l$ bands of extended electrons is therefore exactly 
equivalent to the exact filling of $l$ Landau subbands (with both 
localized and extended states in them) as well as $l$ Landau levels 
in the classical case (i.e., without localization). In such a situation
with the help of (4) we have 
$$
j_x = Nev = ls^E(B)eV(B) = E_y(B)le^2/h \equiv ls(B)e
{E_y(B)\over B}\,\,\,,\eqno(11a)
$$
so that
$$
R_H = {E_y(B)\over j_x} = {B\over ls(B)e} = {h\over
le^2}\,\,\,;\,\,\, B\in (B_a, B_b),\,\,\,\,{\rm say}\,\,.\eqno(11b)
$$
Thus {\it all} the states in $l$ subbands, $ls(B)$, and {\it only
these many} states matter when the Hall effect is quantized irrespective
of the facts that $N$ may be $<$ or even $> ls(B)$.

Returning to $V(B)$ we note that it oscillates about $(D+1)v$.
When $E_F$ is in the $l^{\rm th}$ mobility gap,
$$
V(B) = {ls(B)\pm\eta\over ls^E(B)} v = (D+1)v \pm {\eta\over
ls^E(B)} v\,\,\,;\,\,\,\eta\ge 0,\eqno(12a)
$$
i.e., it decreases from above $(D+1)v$ to below it as $B$
increases. And for $E_F$ lying in the $l^{\rm th}$ band of
extended states, 
$$
V(B) = (D+1)v + {{1\over2}-f\over l-1 + f} Dv \,\,\,;\,\,\, 0 \le
f\le 1,\eqno(12b)
$$
where $f$ is the occupation fraction of the upper most band of
extended states; so $V(B)$ increases from below $(D+1)v$ (for
$f\sim 1$) to above it (for $f\sim 0$) as $B$ increases.
$V(B)=(D+1)v$ for $\eta=0$ and $f={1\over 2}$.

Since $j_x=Nev=n^E(B)eV(B)$, we have
$$
n^E(B) = {N\over D+1\pm \eta/(ls^E(B))} \qquad \mbox{for $E_F$ in
$l^{\rm th}$ mobility gap}\,\,;\eqno(13a)
$$
and
$$
 = {N\over D+1 + {{1\over2} -f\over l-1+f}} \qquad \mbox{for $E_F$ in
$l^{\rm th}$ band of ext.sts.}\,\,.\eqno(13b)
$$
The $n^E(B)$ oscillates about $N/(D+1)$, the value it attains
when $\eta=0$ and $f=1/2$.

To plot $V(B)$ and $n^E(B)$ we make following additional
observations with reference to Fig. 1(c):
\begin{itemize}
\item[(i)] $V(B_a)-V(B_l) = V(B_l)-V(B_b)$\,,\,\,\, since
$$
V(B_a) = [1+D/\{2l(D+1)\}]V(B_l)\,\,\,,{\rm and}\eqno(14a)
$$
$$
V(B_b) = [1-D/\{2l(D+1)\}]V(B_l)\,\,\,.\eqno(14b)
$$
\item[(ii)] $V(B_a) < V(B_c)$ since
\addtocounter{equation}{+4}
\begin{equation}
{V(B_a)\over V(B_c)} = {l-1\over l} {2l(D+1)+D\over
2(l-1)(D+1)+D} < 1\,\,\,.
\end{equation}
\item[(iii)] The number of localized states scanned when $E_F$
moves from its position at $B_a$ to that at $B_l$ is
$s^L(B_a)/2$, and it is $s^L(B_l)/2$ when $E_F$ goes from $B_l$
to $B_b$. Since $s^L(B_l)>s^L(B_a)$ the plateau must be {\it
asymmetric} about the classical $R_H(B)-$ line even under the ideal
conditions of symmetric subbands.
\end{itemize}

The saw-tooth variation of $V(B)$ is shown schematically in Fig.
2(a). The $n^E(B)$ varies in a manner complementary to that of
$V(B)$ --- Fig. 2(b). The bend in each arm of variation is due to
the combined effects of (iii) and (i). The $V(B)$ and $n^E(B)$
will approach finite non-zero values, $(D+1)v$ and $N/(D+1)$
respectively, in the $B\to 0$ limit if $D$ is assumed to be
independent of $B$.

However, $D$ must diverge as $B\to 0$ if $n^E(B)$ must approach
zero in this limit to yield the well known $2d$ localization
result.$^3$ That is, the $(D+1)v-$, and $N/(D+1)-$ lines about
which $V(B)$ and $n^E(B)$ oscillate should indeed stoop upwards
and downwards respectively as shown. In the case of $n^E(B)$ the
$N/(D+1)-$ line can meet the $B-$ axis either at $B=0$ or at a
$B>0$. The former would correspond to the possibility discussed
in the set of references (5) --- $n^E(B)$, on average, would
decrease with $B$, becoming zero only at $B=0$; so the amount of
extra disorder required to convert them into localized states too
would approach zero as $B\to 0$, i.e., $W_c(B)\to 0$ as $B\to
0.^5$ On the other hand if the band-mixing, studied by Haldane
and Yang,$^4$ has to have a noticeable effect to lead to the
floatation of extended states as proposed by Khmelnitskii, and
Laughlin,$^4$ then the $N/(D+1)-$ line should be expected to
converge with decreasing $B$ to a point at $B>0$ --- since the
band-mixing causes the energies of extended states to shift
upwards, as shown by Haldane and Yang,$^4$ with decreasing $B$
besides decreasing in number, the extended states should also be
moving steadily from below the $E_F$ to above it, so their number
below the $E_F$ should deplete faster than in the previous case
and the region below the $E_F$ should become devoid of extended
states well before $B=0$ is reached.

The rate at which the $D$ diverges as $B\to 0$, which we need to
know in order to resolve between the two situations discussed
above, can be determined from the following experiment.

\noindent{\it Suggested experiment:} Within the usual IQHE set up
we propose the following to count the number of occupied
localized and extended states at a given value of $B$ in a sample
of {\it known} $N$. Set $B$ at the value, say $B_a$,
corresponding to the beginning of a pleateau, say $l=2$, and
reduce the number-density of electrons from the initial value
$N$ by changing the gate voltage while keeping the $j_x$ fixed at
$Nev$ and $B$ at $B_a$. This will move the $E_F$ towards the
point $B=B_2 (\equiv B_{l=2})$ of Fig.1c. The quantum Hall
voltage $E_y(B_a)$ will not change in this process but the
classical Hall voltage ${\cal E}_y(B_a) (=B_a/(Ne))$ will increase. By
monitoring the variation of ${\cal E}_y$ the $E_F$ can be moved to the
position corresponding to $B=B_2$ where ${\cal E}_y(B_a)$ will become
equal to $E_y(B_a) (=h/(2e^2))$. Determine the number-density of
electrons at this stage. Suppose it is $N^\prime$. Then
$N^\prime$ will be the number of electrons filling two subbands
{\it exactly} and the electrons removed from the system,
$N-N^\prime$, will be from the localized states. So,
$2(N-N^\prime)$ will be the number of {\it localized} electrons
per subband at $B=B_a$, and we will have
\begin{eqnarray}
S^E(B_a) &=& [N^\prime - 4(N-N^\prime)]/2\,\,\,\,,{\rm and}\\
D(B_a) &=& 4(N-N^\prime)/(5N^\prime-4N)\,\,\,\,.
\end{eqnarray}
The $D(B_2)$ can be similarly determined. The $D(B_b)$ too can be
determined in the above way but by adding the electrons to the
empty localized states in the upper half of the subband $l=2$. In
this way even without knowing the density of states we can
measure $D(B)$ at certain special values of $B$ (such as $B_a,
B_2, B_b, B^2$ etc.) and produce the salient features of the
$n^E(B)$- graph. An experimental set up good enough to produce
sufficiently precise large-$l$ plateaus should enable us to see
whether the $N/(D+1)$-line meets the $B$-axis at $B=0$ or at a $B>0$.

Finally, the answer to question (B) is apparent from the
present analysis --- the number of extended states in a Landau
subband form a vanishing fraction of the total number of states
in it only in the limit $B\to 0$ when $D\to\infty$, otherwise
this ratio is always non-vanishing. This however does not
contradict the possibility of all the extended states occurring
at a single energy in the centre of a subband.$^9$

In summary, simply by splitting $s(B)$ into $s^E(B)$ and $s^L(B)$
and writing $E_y(B)$ in terms of $s^E(B)$ and $V(B)$ we are able
to translate the IQHE result in terms of $V(B)$ and $n^E(B)$
which are found to have novel saw-tooth variations as a function
of $B$. The proposed simple extension of the IQHE experiment to
measure $n^E(B)$ can resolve the controversy about the approach
of the IQHE to the $2d$ localization result in the limit $B\to
0$. The present alternative view of the IQHE result also provides
an easy understanding of the phenomenon.
\\ \\

{\bf Acknowledgements:}
I am grateful to Prof.Sir Sam Edwards for discussions and hospitality 
at the Cavendish Laboratory where this work was initiated. Thanks are 
also due to Prof.D.Shoenberg for discussions. Financial support for 
this work was provided by the Leverhulme Trust, London and the 
Association of Commonwealth Universities.

\newpage
\noindent{\bf References}
\begin{enumerate}
\item K. von Klitzing, G. Dorda and M. Pepper, Phys. Rev. Lett.
	{\bf 45}, 494 (1980).
\item D. Shoenberg, {\it Magnetic Oscillations in Metals}
	(Cambridge University Press, 1984) also pointed out (p. 159) that
	the IQHE result$^1$ should imply a saw-tooth variation of the
	number of mobile electrons with magnetic field but did not give a
	mechanism for it.
\item E. Abraham, P.W. Anderson, D.C. Licciardello and T.V.
	Ramakrishnan, Phys. Rev. Lett. {\bf42}, 673 (1979).
\item D.E. Khmelnitskii, Phys. Lett. {\bf106A}, 182 (1984); R.B.
	Laughlin, Phys. Rev. Lett. {\bf52}, 2304 (1984); K. Yang and R.N.
	Bhatt, {\it ibid} {\bf 76}, 1316 (1996); F.D.M. Haldane and K.
	Yang, {\it ibid} {\bf 78}, 298 (1997) and references therein.
\item D.Z. Liu, X.C. Xie and Q. Niu, Phys. Rev. Lett. {\bf76},
	975 (1996); D.N. Sheng and Z.Y. Weng, {\it ibid} {\bf 78}, 318
	(1997), and references therein.
\item The Hall voltage $E_y(B)=h/(le^2)j_x=B/[ls(B)e]j_x$ {\it
	along the $l^{\rm th}$ plateau} with the system current $j_x$
	fixed at $Nev(N=$ no. density of electrons in the system;
	$v=$drift velocity). Note that exactly $ls(B)$ states matter
	eventhough $N$ may be $>$ or even $< ls(B)$.
\item R.B. Laughlin, Phys. Rev. B{\bf23}, 5632 (1981).
\item The $E_y$ is typically of order 10$^{-5}$;
	$s^E(B)=s(B)/(D+1)$ where $s(B)\sim10^9$ per mm for $B\sim 10$ T.
\item See e.g., Y. Huo and R.N. Bhatt, Phys. Rev. Lett. {\bf68},
	1375 (1992).
\end{enumerate}

\newpage
{\bf Figure Captions:}

Figure.1:\\
\indent (a) The integer quantum Hall effect geometry; (b) the IQHE 
plateaus in the Hall voltage (for a fixed system current $j_{x}$); 
broken line shows the classical Hall effect result; (c) density of 
states (DOS) comprising disorder-broadened Landau levels with extended 
states in the middle and localized states in the shaded regions; (d) 
enlargement of a portion of (c) --- the cross-hatched region has $\eta$ 
localized states.

Figure.2:\\
\indent Schematic representation of the saw-tooth variation of (a) 
drift velocity $V(B)$, and (b) number of extended electrons $n^E(B)$.
The oscillations happen in (a) and (b) respectively about $(D+1)v-$ 
and $N/(D+1)-$ lines where $\cdots$ and --- represent $D\to\infty$ 
without and with band mixing.

\end{document}